# Analisis Fungsi Keamanan Terhadap Kinerja Router Pada Jaringan Berkecepatan Tinggi
*Security Function Analysis on Performance of High Speed Router Networking*

**Kurniawan Dwi Irianto**
*Jurusan Teknik Informatika, Fakultas Teknologi Industri, Universitas Islam Indonesia*
Jl. Kaliurang KM 14,5 Sleman Yogyakarta 55584, Telp: (0274) 895287
E-mail: k.d.irianto@uii.ac.id

**Abstrak**

Router adalah sebuah perangkat pada jaringan komputer yang digunakan untuk meneruskan paket data. Karena fungsinya, router menjadi peran penting dalam mengukur kinerja jaringan komputer. Router perlu memproses paket data yang masuk dengan kecepatan tinggi dan delay sedikit mungkin. Oleh karena itu, router harus didukung dengan sistem keamanan yang baik juga. Aktivasi aplikasi keamanan dan enkripsi data pada router dan firewall dapat mengkonsumsi sumber daya CPU untuk komputasi tambahan dan hal ini dapat mempengaruhi penurunan kinerja router. Selain itu, mekanisme keamanan di router bisa membawa kemungkinan kemacetan di jaringan pada saat ON/OFF fungsi keamanan sehingga dapat mempengaruhi kinerja router dalam meneruskan paket data. Sebuah analisis kuantitatif didasarkan pada teori antrian dapat digunakan untuk mengukur dan memprediksi serta meningkatkan kinerja router. Dalam tulisan ini, beberapa skenario simulasi akan dilakukan untuk menganalisis pengaruh fungsi keamanan di router pada jaringan kecepatan tinggi. Skenario tersebut menggunakan distribusi Generalized Exponential (GE-type) untuk mencerminkan interarrival bursty dan waktu servis pada router. Hasil penelitian menunjukkan bahwa fungsi keamanan di router bisa membawa penurunan pada kinerjanya.

**Kata kunci**— router, keamanan, kinerja, teori antrian, jaringan berkecepatan tinggi

*Abstract*

*Router is a device in the computer networks that is used to forward a data packet. Because of its function, router becomes an important role in measuring the performance of computer networks. It needs to process the incoming data packets with the high speed and minimum delay. Therefore, router must have supported with a good security system as well. The activation of security applications and data encryptions in the router and firewall can consume the resources of CPU from the addition computations and they are able to affect the degradation of its performance. Furthermore, the security mechanisms in the router could bring a possibility of congestion in the networks during the ON/OFF functioning of security so it might influence the performance of router in forwarding the data packets. A quantitative analysis based on queuing theory may be used to measure and predict as well as improve the performance of router. In this paper, some scenarios of simulation will be carried out to analyze the effect of security function in the router with high speed networks. The schemes use generalized exponential (GE-type) distribution to reflect bursty interarrival and service time at a router. The results show that the security function in the router could bring the decreasing of its performance.*

*Keywords*— router, security, performace, queueing theory, high speed networks

## 1. PENDAHULUAN

*Router* adalah sebuah alat yang berfungsi untuk meneruskan paket data antar jaringan komputer. Hal tersebut menjadikan *router* sebagai salah satu kunci penting dalam kinerja sebuah jaringan komputer karena *router* dituntut untuk memproses paket data yang masuk dengan kecepatan tinggi dan *minimum delay* [1]. Karena peranan *router* yang sangat penting maka *router* harus didukung dengan sistem keamanan yang tinggi pula. Akan tetapi hal tersebut dapat memberikan dampak buruk terhadap kinerja jaringan komputer sebab pengaktifan fungsi-fungsi keamanan yang mengharuskan *router* memproses data tambahan yaitu berupa data asli dan data tambahan yang dimasukkan oleh aplikasi keamanan [2]. Pengaktifan aplikasi-aplikasi keamanan yang berjalan pada *router* dan *firewall* serta aplikasi yang menyedikan enskripsi data dapat mengkonsumsi penggunaan CPU. Hal itu disebabkan karena adanya komputasi tambahan dan menyebabkan terjadinya penurunan kinerja. Disamping itu, mekanisme keamanan dapat menjadikan *router* berpotensi sebagai sebab terjadinya kemacetan pada jaringan dimana kecepatan CPU berubah-ubah dari aktif "ON" kemudian non-aktif "OFF" sehingga dapat memberikan dampak pada penurunan kecepatan *router* dalam meneruskan paket data [3].

Untuk mendapatkan kinerja yang optimal dari sebuah sistem, diperlukan tahapan-tahapan perancangan seperti pemodelan, estimasi dan analisis parameter serta *output* yang akan dihasilkan oleh sistem. Sistem yang dirancang dalam pola tersebut memiliki beberapa keuntungan, salah satunya adalah dapat mengetahui tantangan apa saja yang akan dihadapi yang disebabkan oleh pemrosesan, padatnya lalu lintas data dan kebutuhan keamanan dalam sistem [4].

Oleh sebab itu, analisis kuantitatif berdasarkan teori antrian dapat digunakan pada sistem tersebut untuk mengukur, memprediksi serta meningkatkan kinerja *router* sebuah jaringan [5]. Meskipun model analitik dari teori antrian menyediakan solusi yang menarik, akan tetapi seiring dengan kompleksitas sistem yang terus meningkat maka dibutuhkan penyederhanaan asumsi untuk mendapatkan persamaan matematis yang sesuai dan komponen-komponen kinerja yang diharapkan.

*1.1. Mekanisme Keamanan Router*

*Router* dan *firewall* merupakan komponen yang sangat penting dalam jaringan komputer kerena mereka harus bisa melindungi paket data yang terkirim dan hanya mengijinkan *user* terotentikasi yang bisa mengirimkan data [6]. Disamping itu, *router* dan *firewall* juga melakukan penyaringan lalu lintas data yang masuk-keluar jaringan berdasarkan dengan kriteria protokol tertentu untuk melindungi komponen jaringan. Hal ini bisa dicapai dengan menguji *header* paket-paket data dan melakukan rute ulang lalu lintas data [3].

Yang perlu diperhatikan bahwa penyaringan data dalam *firewall* dapat ditambah atau dikurangi sesuai dengan beberapa kondisi yang dibutuhkan. Sejak semua fungsionalitas keamanan bekerja pada *router* atau *firewall* maka hal ini memberikan pengaruh terhadap layanan keamanan. Fungsi ketersedian, keutuhan dan kerahasian jaringan menjadikan *router* sebagai komponen yang tingkat keamanannya tinggi dan merupakan titik yang esensial [6].

*1.2. Aplikasi Keamanan ACL*

Sebagaimana yang telah dijelaskan sebelumnya bahwa tugas *router* dan *firewall* adalah untuk menyaring lalu lintas paket-paket data yang masuk dengan mengijinkan beberapa paket masuk dan menolak paket yang lain sesuai dengan kriteria tertentu pada *header* paket. Meskipun demikian, tetap terdapat kemungkinan terjadinya serangan pada *router* dan *firewall* seperti *Denial of Service* (DoS), *session hijacking*, *unauthorized access*, *rerouting* dst. Aplikasi keamanan ACL dapat mendeteksi dan menjaga beberapa akses serangan pada *router* dan *firewall* seperti *Address Resolution Protocol* (ARP) *Spoofing Attack* dan serangan DoS [6].

Fungsionalitas ACL dapat dilihat pada gambar 1. Aplikasi ACL yang berjalan pada *router* dalam jaringan yang kompleks dapat meningkatkan efisiensi keamanan mereka [7]. Akan tetapi sebagaimana yang telah disinggung sebelumnya bahwa aktifasi ACL mengkonsumsi penggunaan CPU *router*.

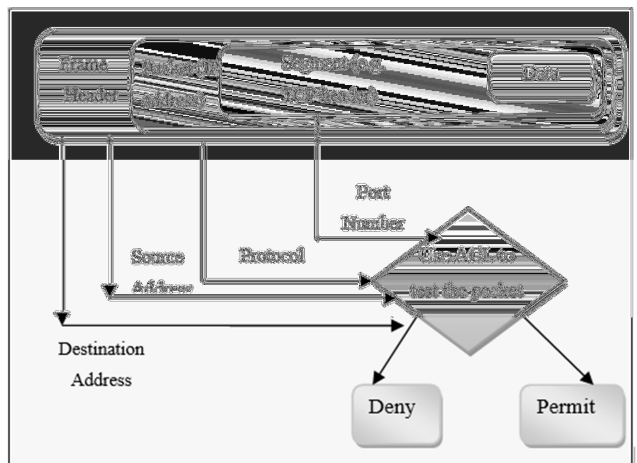

Gambar 1. Konsep Fungsi ACL Pada Router [7]

*1.3. Model Jaringan Antrian*

Teori antrian adalah sebuah ilmu dalam mempelajari fonomena menunggu dalam antrian dan teori ini berdasarkan pada toeri probabilitas terapan. Tujuan dari teori ini adalah untuk untuk mengukur kinerja dari sebuah sistem [8]. Sebagai ilustrasi, sebuah model sederhana dari sistem antrian dapat dipresentasikan oleh sebuah antrian *multi server* yang biasanya dinotasikan sebagai M/M/c/N dengan c (C>1), kapasitas terbatas N (N>0) atau kapasitas tak terbatas (N --> + infinity) dan *interarrival* dan waktu servis eksponensial. Skema dari sistem antrian sederhana dapat dilihat pada gambar 2.

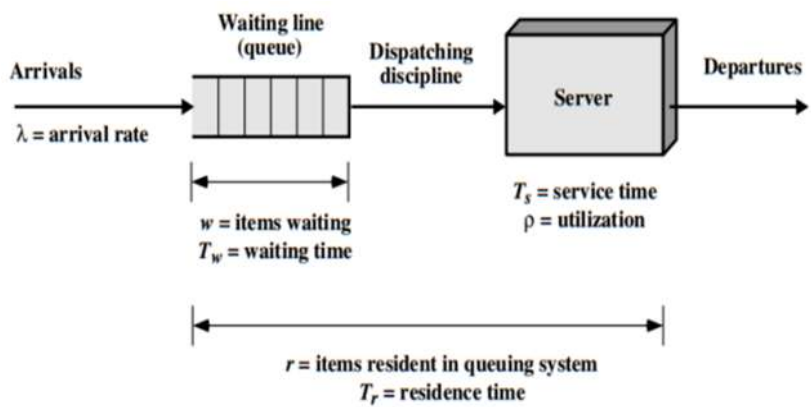

Gambar 2. Model Antrian Sederhana [8]

## 2. METODE

Pada penelitian ini, fungsi ACL dan *forwarding* merepresentasikan sebuah *router* dengan atau tanpa mekanisme keamanan ACL dan itu bisa dimodelkan oleh antrian GE/GE/1/N/FCFS dan GE/GE/1/N/HOL [9]. Model antrian ini juga bisa diterapkan untuk *firewall* karena *firewall* memiliki cara kerja yang mirip dengan *router*.

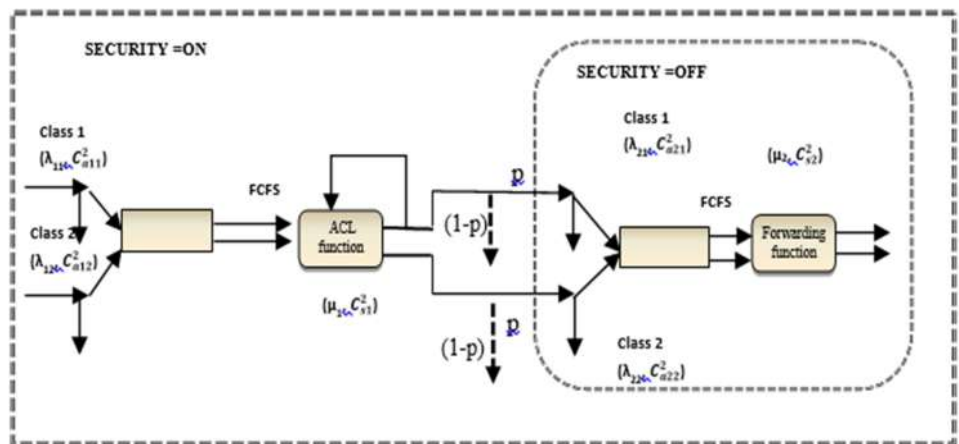

Gambar 3. Model Antrian Secara Umum Pada Router

Ketika paket data tiba di *router*, paket akan diuji pertama kali oleh layanan kemanan ACL pada *node* 1. Jika *node* fungsi ACL penuh, paket tersebut akan hilang. Bila tidak, setelah servis selesai, paket ini bisa diterima untuk transmisi oleh *router* dengan probabilitas p atau paket tersebut ditolak dengan probabilitas 1-p. Jika paket yang diterima menemukan antrian dari aliran data pada node 2 dengan fungsi *forwarding* saat kapasitas penuh, itu akan diblok sesuai dengan mekanisme bloking yang diterapkan.

Adapun metodologi yang digunakan dalam mengembangkan model analitik adalah dengan menggunakan konsep teori antrian. Konsep ini sangat penting dan bermanfaat dalam mengukur, menilai dan mengembangkan sebuah sistem khususnya dalam bidang jaringan komputer. Adapun pun program simulasi akan dibangun dengan menggunakan konsep *Discrete Event Simulation* (DES). DES adalah suatu model dalam sistem yang ditunjukkan sebagai kondisi variabel yang berubah seketika pada titik waktu (*point in time*) yang terpisah. (dalam persamaan matematika dapat dikatakan sistem berubah dalam titik waktu tertentu secara *countable*). Dalam titik waktu ini akan terjadi suatu *event*, dimana *event* didefinisikan sebagai suatu kejadian yang dapat mengubah kondisi suatu sistem. Program simulasi dengan DES disusun menggunakan pemograman Java dan *tool* yang dipakai adalah Netbeans 8.0.2.

*2.1. Prosedur Pengujian*

Simulasi dilakukan dengan konsep DES dan bantuan pustaka-pustaka pada pemograman JAVA dalam memodelkan GE/GE/C/N/FCFS dan GE/GE/C/N/HOL. Hal ini dimaksud untuk mengukur kinerja *router* dalam hal; rerata waktu respons (*mean response time*), rerata panjang antrian *(mean queue length)*, utilisasi (*utilization*) dan jumlah paket data yang hilang (*packet loss*). Program dijalankan sebanyak 20 kali iterasi dan menggunakan distribusi tipe GE (*General Exponential*). Tabel 1 menunjukan beberapa skenario simulasi yang dilakukan pada sistem/model gambar 4 dan 5.

Diasumsikan bahwa pada *router* terdapat lebih dari satu CPU (CPU tunggal dan CPU Quad-Core) yang bertugas untuk melakukan proses komputasi fungsi keamanan, dalam hal ini adalah ACL. Proses kerja sistem antrian menggunakan dua pola, yaitu FCFS (*First Come First Served*) dan PQ (*Priority Queue*) / HOL (*Head of the Line*). Cara kerja FCFS adalah setiap paket data yang datang paling duluan, itulah yang akan diproses terlebih dahulu dan setiap paket data akan dilayani sesuai dengan urutan antrian berdasarkan waktu kedatangan. Sementara pada HOL (Head of the Line), setiap paket data yang datang akan dilayani sesuai dengan waktu kedatangan dan prioritas. Jika terdapat dua paket data yang datang secara bersamaan maka paket data yang memiliki prioritas lebih tinggi yang akan dilayani terlebih dahulu. Atau jika ada paket data dengan prioritas lebih tinggi datang sementara paket data yang lain dengan prioritas lebih rendah sedang diproses pada CPU maka proses tersebut harus dihentikan dan paket data lebih rendah masuk ke antrian untuk membiarkan paket data lebih tinggi diproses oleh CPU. Sementara itu, SCV (*Squad Coefficient of Variation*) merepresentasikan kepadatan lalu lintas paket data yang datang. Semakin tinggi nilai SCV, semakin tinggi pula tingkat padatnya lalu lintas data.

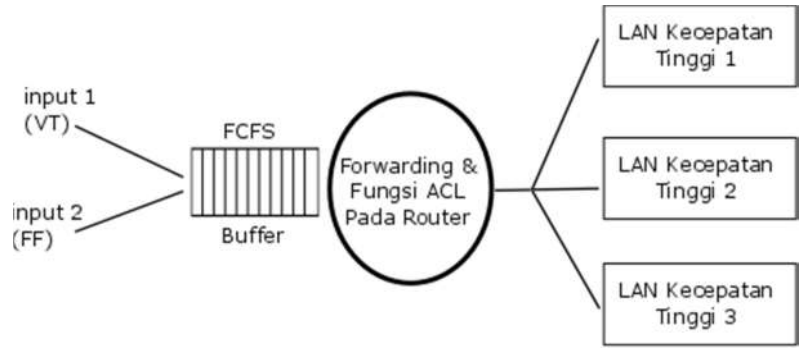

Gambar 4. Perancangan Model Antrian dengan CPU Tunggal pada Router

Diasumsikan pula, terdapat 2 input paket data yang masuk kedalam *router* dan mempunyai prioritas yang berbeda, yaitu Video Streaming (VT) dan Transfer File (FF). Input yang pertama membutuhkan *bandwidth* yang tinggi, sensitif terhadap *router delay* dan membutuhkan aplikasi yang *realtime*.

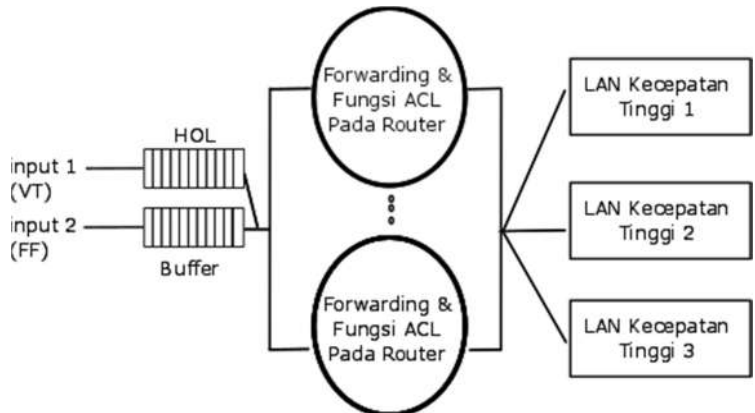

Gambar 5. Perancangan Model Antrian Dengan CPU Lebih Dari Satu Pada Router

Untuk input prioritas tinggi memiliki rata-rata waktu datang $\lambda_1$ sebesar $1 \times 10^5$ sampai $10 \times 10^5$ paket/detik dan $\lambda_2 = 5 \times 10^5$ paket/detik untuk input 2 yang berprioritas rendah. SCVa1=SCVa2=4 untuk input 1 dan 2. Adapun MU adalah *Mean Service Rate* atau rata-rata waktu yang dibutuhkan CPU dalam memproses sebuah paket data. Rata-rata waktu servis (MU) $\mu = 17 \times 10^5$ paket/detik. SCVs2= 4 untuk MU. CPU mempunyai keterbatasan kapasitas dalam menampung antrian paket data yang disebut dengan *buffer*. Untuk itu, ukuran buffer dalam percobaan ini diset sebanyak 50 (N=50) untuk masing-masing FCFS dan HOL. Ini berarti CPU hanya bisa menampung 50 paket data yang datang. Bila kapasitas *buffer* penuh, maka paket data yang datang tidak akan bisa masuk ke dalam antrian dan dibuang. Fitur keamanan diset aktif (ON) dan non-aktif (OFF). Hal ini bertujuan untuk melihat perubahan yang terjadi pada kinerja CPU. Percobaan dilakukan dengan menerapkan 4 skenario sebagaimana yang terdapat pada tabel 1.

Tabel 1: Beberapa Skenario Pada Simulasi

| Skenario | Kriteria | Deskripsi |
|---|---|---|
| A | FCFS & PQ; SEC = OFF, c = 4; N = 50; SCV = 4; MU=17. | Membandingkan FCFS dan PQ pada kinerja router |
| B | PQ; SEC= ON, OFF; c=4; N=50; MU=17; SCVa1=SCVa2=5; PQ; SEC= ON,OFF; c=4; N=50; MU=17; SCVa1=SCVa2=10; | Mengukur pengaruh SCV pada kinerja router terhadap lalu lintas paket-paket yang masuk |
| C | PQ; SEC=OFF; c=1; c=4; N=50; MU=17; SCV=4. | Melihat efek dari CPU tunggal dan Multi CPU pada router |
| D | PQ; SEC=OFF, ON; c=4; N=50; MU=17; SCV=4. | Mengamati kinerja router saat fitur keamanan aktif dan non-aktif |

PQ= Priority Queue, FCFS= First Come First Served, SEC= Security, c = jumlah CPU pada router, N= kapasitas router, MU= rata-rata waktu service router, SCV= jumlah Squared Coefficient of Variation.

## 3. HASIL DAN PEMBAHASAN

*3.1. Skenario A: Membandingkan FCFS dan PQ/HOL Pada Kinerja Router*

Skenario A bertujuan untuk melihat seberapa besar pengaruh perbedaan antara model antrian FCFS (First Come First Served) dan PQ (Priority Queue) pada sistem ini. Untuk FCFS, setiap paket data yang masuk ke *router* akan dilayani berdasarkan waktu kedatangan. Paket yang datang duluan akan dilayani terlebih dahulu sementara paket data yang datang belakangan akan dilayani paling akhir. Jika terdapat sebuah paket data baru yang masuk ke *router* kemudian disaat yang bersamaan *router* sedang melayani paket data yang lain maka paket data yang masuk tersebut harus antri ke dalam buffer untuk menunggu giliran agar bisa dilayani oleh *router*. Sementara itu pada PQ, setiap paket data yang masuk ke dalam *router* akan dilayani berdasarkan dengan prioritas yang dimiliki. Jika prioritasnya lebih tinggi maka akan dilayani terlebih dahulu meskipun waktu kedatangannya lebih lambat daripada paket data yang prioritasnya lebih rendah. Jika sebuah paket data A (dengan prioritas lebih tinggi) masuk ke dalam *router* sementara *router* sedang melayani paket data B (dengan prioritas lebih rendah) maka *router* harus menghentikan proses B untuk sementara dan segera melayani paket data A. Pada skenario ini, semua fungsi keamanan tidak diaktifkan dan input 1 (VT) memiliki prioritas yang lebih tinggi daripada input 2 (FF).

Gambar 6 menunjukkan hasil simulasi untuk *Mean Response Time* (rerata waktu respon) pada skenario A. *Mean Response Time* (MRT) adalah waktu rerata yang dibutuhkan sebuah paket data selama berada di dalam sistem, mulai dari masuk ke dalam *router* hingga keluar dari *router*. W1 adalah MRT untuk input 1 (VT) dalam hal ini adalah paket data untuk video streaming, sementara W2 untuk input 2 (FF) yaitu paket data file transfer. Perlu diketahui bahwa MRT akan semakin baik jika nilainya semakin sedikit, artinya paket data yang masuk ke *router* hanya membutuhkan waktu yang sedikit. Dari gambar 6 dapat diketahui bahwa nilai W1-PQ-Sec=OFF lebih rendah dari pada W1-FCFS-Sec=OFF. Ini menandakan bahwa kinerja model antrian untuk paket data video streaming yang masuk ke *router* lebih baik menggunakan PQ daripada FCFS. Demikian pula dengan nilai W2, dapat dilihat bahwa W2-PQ-Sec=OFF mempunyai nilai yang lebih sedikit dibandingkan dengan W2-FCFS-Sec=OFF. Hal ini juga sebagai bukti bahwa model antrian PQ lebih unggul daripada FCFS.

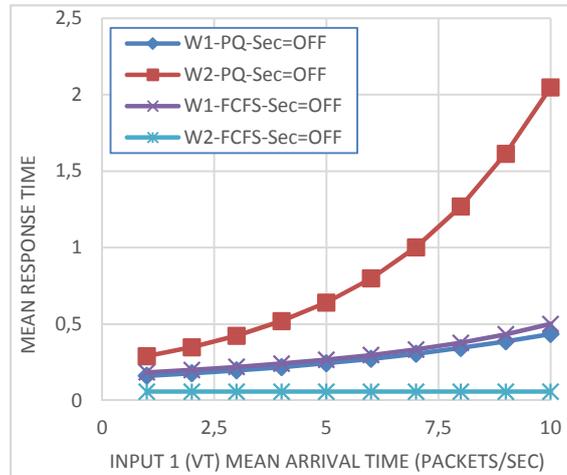

Gambar 6. Mean Response Time (Rerata Waktu Respone) Pada Skenario A

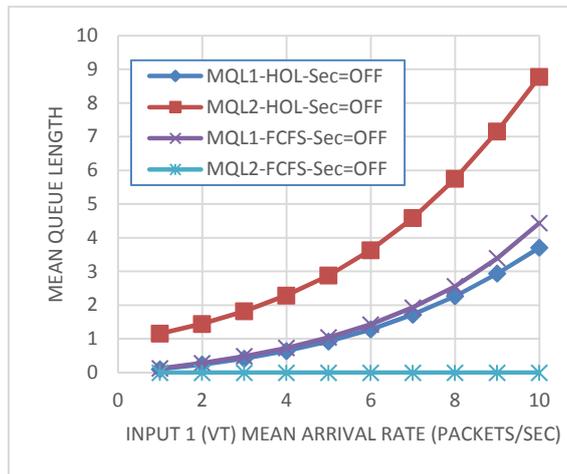

Gambar 7. Mean Queue Length (Rerata Panjang Antrian) Pada Skenario A

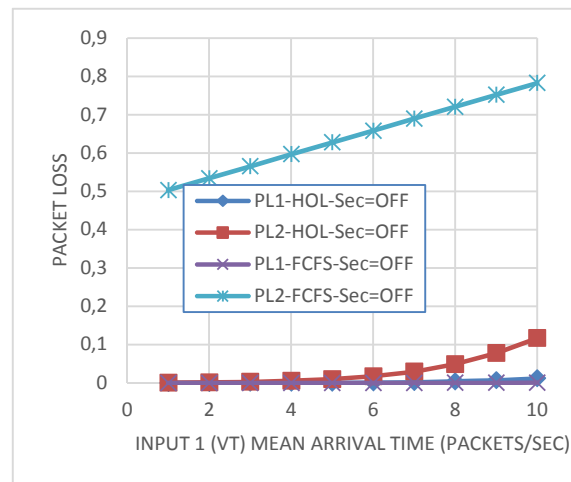

Gambar 8. Packet Loss (Paket Data Yang Hilang) Pada Skenario A Adapun gambar 7-9 menunjukkan nilai untuk mean queue length (rerata waktu antrian), packet loss (paket data yang hilang) dan total utilization (utilisasi total) pada skenario A. Yang dimaksud dengan HOL (Head of the Line) adalah sama dengan PQ (Priority Queue). Dari gambar 7-9, dapat kita lihat bahwa hasil pemodelan dengan HOL lebih baik dibandingkan dengan FCFS. Untuk MQL (mean queue length) dan PL (Packet Loss) akan semaik baik jika nilainya semakin kecil dan SU (total utilization) akan optimal jika nilainya semakin besar. Kita dapat lihat dan bandingkan nilai antara MQL1-HOL-Sec=OFF dan MQL1-FCFS-Sec=OFF untuk rerata panjang antrian, PL1-HOL-Sec=OFF dan PL1-FCFS-Sec=OFF untuk data paket yang hilang, TU1-HOL-Sec=OFF dan SU1-FCFS-Sec=OFF untuk utilisasi total.

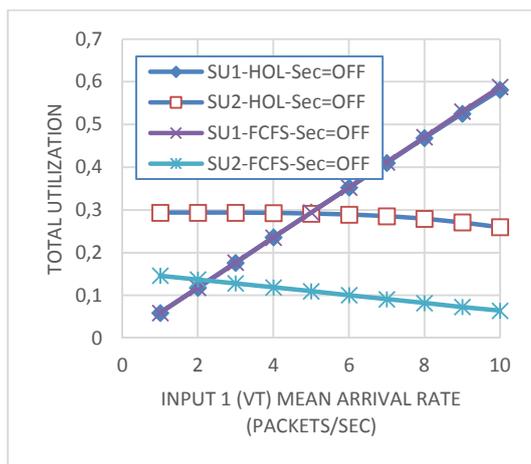

Gambar 9. Total Utilization (Utilisasi Total) Pada Skenario A

*3.2. Skenario B: Mengukur Pengaruh SCV Pada Kinerja Router Terhadap Lalu Lintas Paket-Paket Yang Masuk*

Pada skenario ini, nilai SCV dirubah dan dibandingkan untuk melihat efek dari lalu lintas paket data yang masuk dan melawati *router*. Gambar 10-13 menampilkan hasil simulasi pada skenario ini.

Untuk rerata waktu respon bisa dilihat pada gambar 10. Dari gambar tersebut dapat diketahui nilai rerata waktu respon lebih baik ketika nilai SCV semakin kecil. Hal ini menunjukkan bahwa sistem akan bekerja secara baik ketika laju lalu lintas data yang masuk ke router semakin sedikit. Jika lalulintas data yang masuk ke *router* semakin tinggi, maka sistem harus bekerja lebih keras untuk melayani paket data tersebut yang menyebabkan rerata waktu responnya semakin besar atau tinggi.

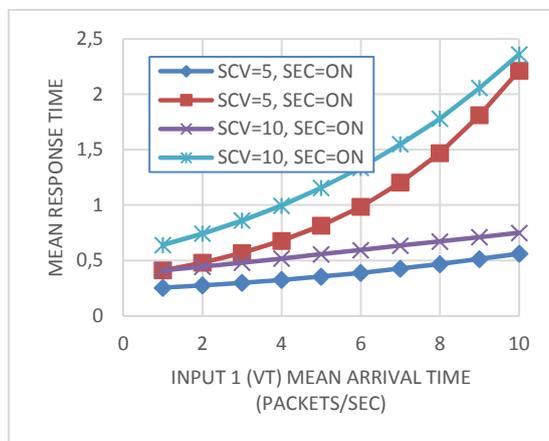

Gambar 10. Mean Response Time (Rerata Waktu Respone) Pada Skenario B

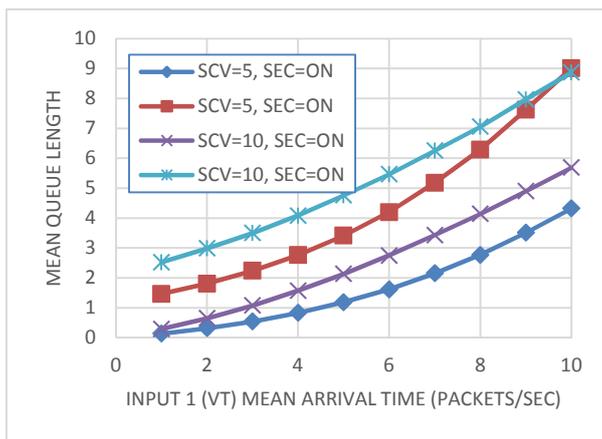

Gambar 11. Mean Queue Length (Rerata Panjang Antrian) Pada Skenario B

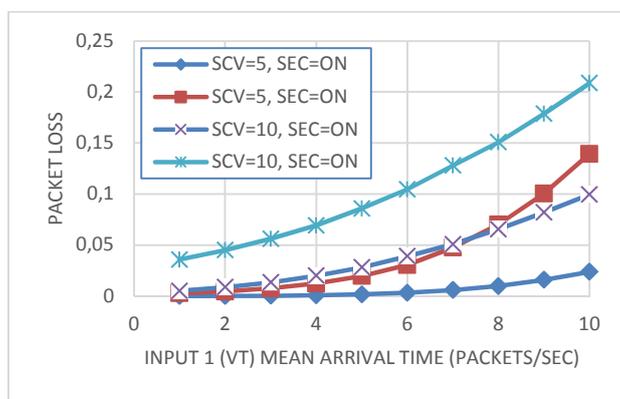

Gambar 12. Packet Loss (Paket Data Yang Hilang) Pada Skenario B

Demikian pula pada gambar 11-13. Dapat dilihat bahwa nilai rerata panjang antrian, paket data yang hilang dan utilisasi total akan baik ketika nilai SCV semakin kecil atau laju lalulintas data semakin sedikit. Bisa kita bandingkan nilai "SCV=5, Sec=ON Input1" dan "SCV=10, Sec=ON Input1".

Dari hasil simulasi pada skenario B, dapat kita simpulkan bahwa SCV juga dapat mempengaruhi kinerja dari sebuah *router* dimana nilai SCV merepresentasikan laju kepadatan lalu lintas data yang masuk ke dalam *router*.

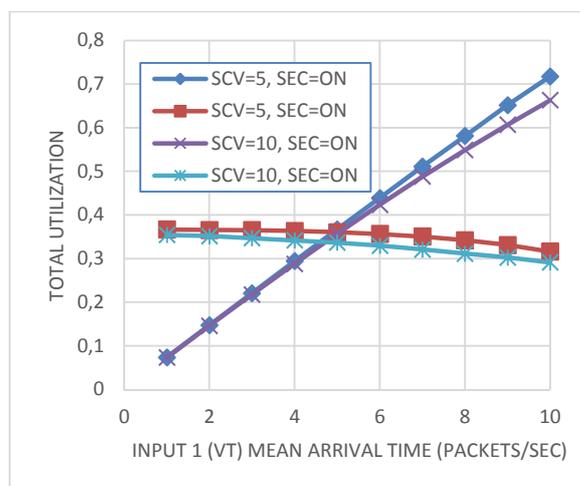

Gambar 13. Total Utilization (Utilisasi Total) Pada Skenario B

*3.3. Skenario C: Melihat Efek Dari CPU Tunggal Dan Multi CPU Pada Router*

Gambar 14-17 menunjukkan hasil simulasi untuk skenario C. Skenario ini bertujuan untuk melihat perbedaan antara *router* yang menggunakan CPU tunggal dan yang menggunakan multi CPU. Dari hasil simulasi yang ditampilkan, dipastikan bahwa *router* dengan multi CPU mempunyai kinerja yang lebih baik dibandingkan dengan *router* yang menggunakan CPU tunggal. Nilai rerata waktu respon semakin baik ketika menggunakan *router* multi CPU, begitu pula untuk nilai rerata panjang antrian, paket data yang hilang dan utilasasi total. Ini berarti semakin banyak jumlah CPU yang dimiliki sebuah *router*, maka semakin baik kinerja dari *router* tersebut dalam melayani semua paket data yang masuk.

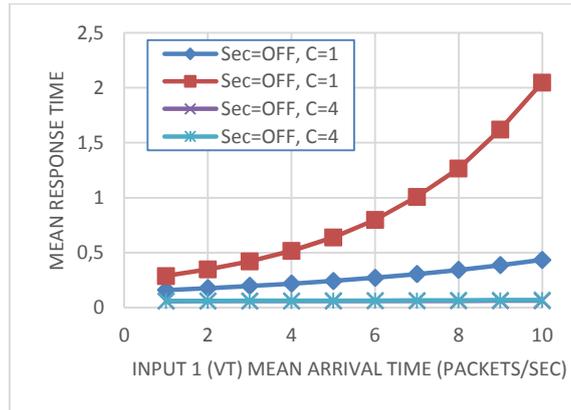

Gambar 14. Mean Response Time (Rerata Waktu Respone) Pada Skenario C

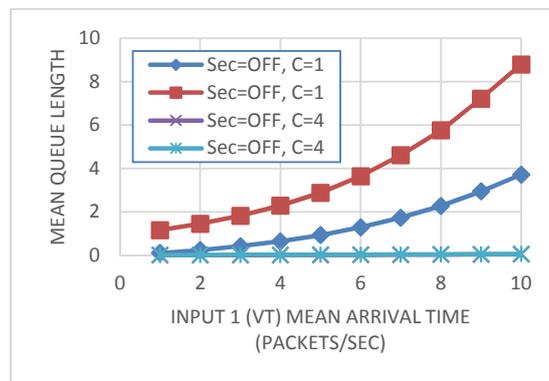

Gambar 15. Mean Queue Length (Rerata Panjang Antrian) Pada Skenario C

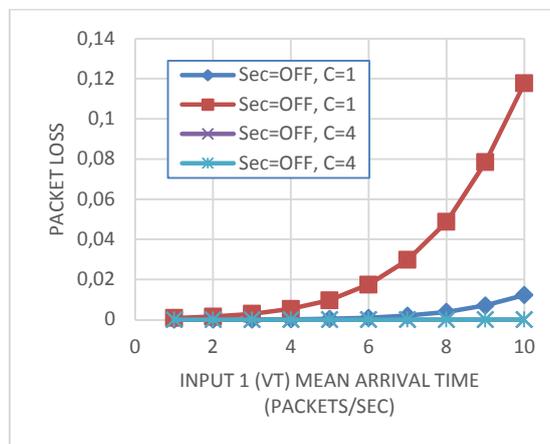

Gambar 16. Packet Loss (Paket Data Yang Hilang) Pada Skenario C

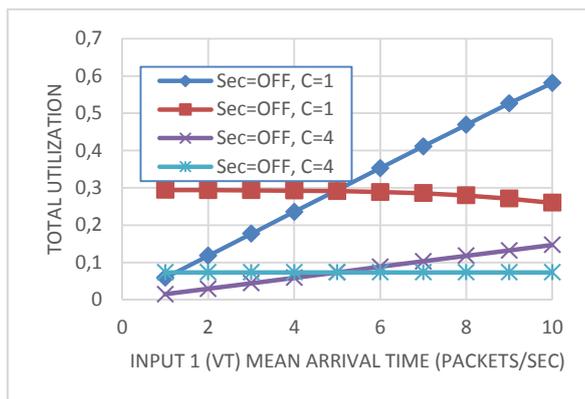

Gambar 17. Total Utilization (Utilisasi Total) Pada Skenario C

*3.4. Skenario D: Mengamati Kinerja Router Saat Fitur Keamanan Aktif Dan Non-Aktif*

Pada skenario D, fitur keamanan diaktifkan dan dinon-aktifkan. Hal ini bertujuan untuk mengukur seberapa besar efek fitur keamanan bagi kinerja *router*. Hasil simulasi pada skenario D dapat dilihat pada gambar 18-21. Pada skenario ini, terdapat 4 CPU pada *router* dan fitur keamanan diaktifkan dan dinon-aktifkan. Untuk rerata waktu respon (gambar 18), *router* mencapai kinjerja yang maksimal ketika fitur keamanannya dinon-aktifkan. Hal ini dikarenakan *router* tidak perlu lagi melakukan komputasi tambahan untuk mengaktifkan proses enskripsi data sehingga semua paket data yang masuk dapat diproses dengan cepat. Demikian juga hasil yang ditunjukkan pada gambar 19 memperlihatkan bahwa rerata panjang antrian yang paling sedikit ketika fitur keamanan dinon-aktifkan.

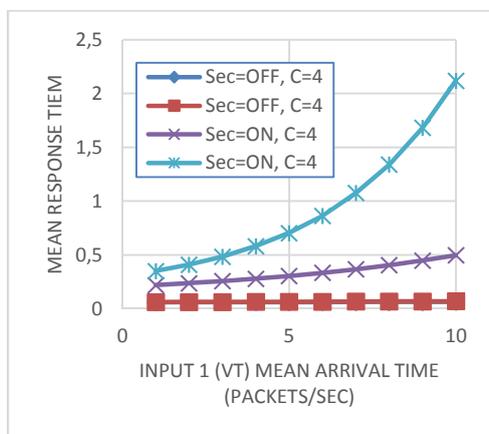

Gambar 18. Mean Response Time (Rerata Waktu Respone) Pada Skenario D

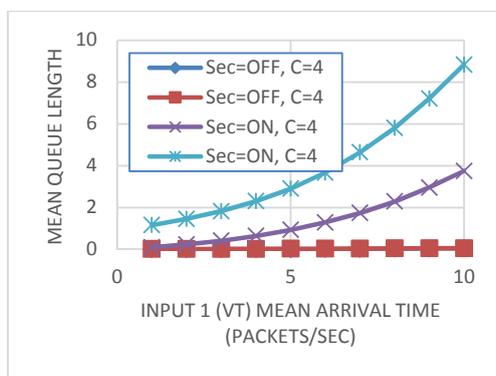

Gambar 19. Mean Queue Length (Rerata Panjang Antrian) Pada Skenario D

Hasil yang sama juga diperlihatkan pada gambar 20-21 untuk nilai paket data yang hilang dan utilisasi total. Data paket yang hilang paling sedikit ketika fitur keamanan dinon-aktifkan dan utilisasi total bisa lebih maksimal.

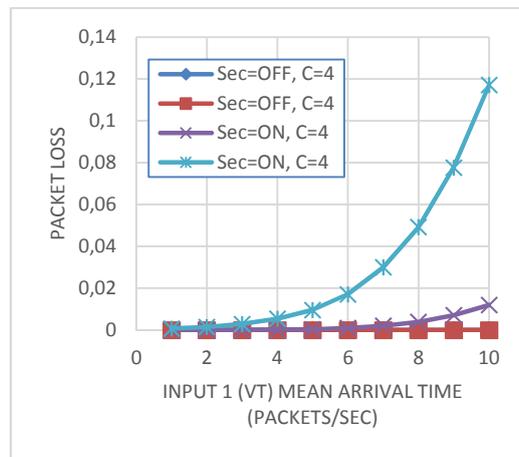

Gambar 20. Packet Loss (Paket Data Yang Hilang) Pada Skenario D

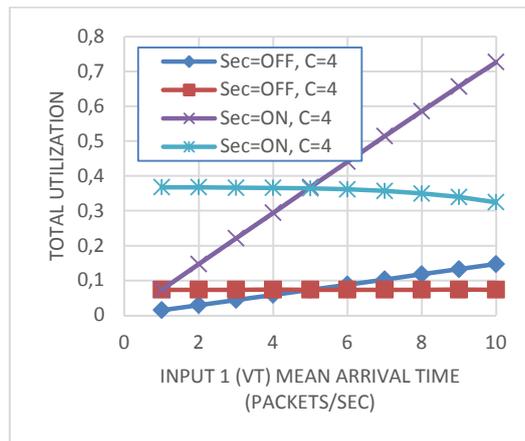

Gambar 21. Total Utilization (Utilisasi Total) Pada Skenario D

## 4. KESIMPULAN

Dari hasil pengujian dan simulasi yang dilakukan pada skenario A-D dan sebagaimana juga yang sudah dibahas maka dapat kita simpulkan bahwa fitur kemanan pada router dapat membawa pengaruh negatif pada kinerja router. Hal tersebut dapat dilihat dari hasil rerata waktu respon, rerata panjang antrian, data paket yang hilang dan utilisasi total. *Router* dengan fitur keamanan dinon-aktifkan memiliki kinerja yang lebih baik ketika fitur keamanan diaktifkan. Hal ini menunjukkan bahwa fitur keamanan membawa dampak negatif pada *router* sehingga menurunkan kinerja *router*. Kemudian dapat disimpulkan juga bahwa *router* dengan multi CPU memberikan kinerja yang lebih baik dari pada *router* dengan CPU tunggal serta laju kepadatan lalu lintas data yang masuk ke *router* juga dapat mempengaruhi kinerja *router*. Dan terakhir, pemodelan dengan PQ/HOL mempunyai kinerja lebih baik daripada FCFS pada *router*.

## UCAPAN TERIMA KASIH

**DAFTAR PUSTAKA**